\documentclass[10pt]{iopart}

\usepackage{iopams, color, enumerate, epsfig, graphicx}
\graphicspath{{graph/}}

\newcommand{\et}{\widetilde{e}}
\newcommand{\half}{\frac{1}{2}}

\newcommand{\Hloop}{H_{\rm loop}}
\newcommand{\Hmg}{H_{\rm MG}}
\newcommand{\Hrsos}{H_{\rm RSOS}}
\newcommand{\Hxxz}{H_{\rm XXZ}}
\newcommand{\mt}{\widetilde{m}}
\newcommand{\On}{{\rm O}(n)}
\newcommand{\Rc}{\check{R}}
\newcommand{\SU}{{\rm SU}}
\newcommand{\thetaint}{\theta_{\rm int}}
\newcommand{\Uq}{{\rm U}_q[\SU(2)]}
\newcommand{\Zbb}{\mathbb{Z}}

\begin{document}

\title{A Temperley-Lieb quantum chain with two- and three-site interactions}

\author{Y. Ikhlef$^1$, J.L. Jacobsen$^2$ and H. Saleur$^{3,4}$}

\address{$^1$ Rudolf Peierls Centre for Theoretical Physics, 1 Keble Road, Oxford OX1~3NP, United Kingdom}
\address{$^2$ Laboratoire de Physique Th\'eorique de l'Ecole Normale Sup\'erieure, 24 rue Lhomond,
  75231 Paris, France}
\address{$^3$ Institut de Physique Th\'eorique CEA, IPhT, 91191 Gif-sur-Yvette, France}
\address{$^4$ Department of Physics, University of Southern California, Los Angeles CA 90089-0484, USA}

\eads{\mailto{y.ikhlef1@physics.ox.ac.uk}}

\maketitle

\begin{abstract}
  We study the phase diagram of a quantum chain of spin-$1/2$ particles whose world lines
  form a dense loop gas with loop weight $n$. In addition to the usual two-site interaction
  corresponding to the XXZ spin chain, we introduce a three-site interaction. The resulting
  model contains a Majumdar-Ghosh-like gapped phase and a new integrable point, which we solve
  exactly. We also locate a critical line realizing dilute $\On$ criticality, without introducing
  explicit dilution in the loops. Our results have implications for anisotropic spin chains, as
  well as anyonic quantum chains.
\end{abstract}
\pacs{05.50.+q,11.25.Hf,02.10.Ox,05.30.Pr}

\section{Introduction}

The study of lattice models of self-avoiding loops is a paradigm
within two-dimensional (2D) statistical physics. In these models, the
Boltzmann weights for loop configurations are encoded
in an algebra. A simple example is the dense loop 
model on the square lattice, associated to the Temperley-Lieb (TL)
algebra~\cite{TL71}, and for which critical points are present for 
a loop weight $n \in [-2,2]$.
This algebra is also realized by several 2D statistical
models, such as the Ising and Potts model, the Restricted Solid-On-Solid (RSOS) model~\cite{RSOS}
or the six-vertex (6V) model. 
At their integrable points and in the very anisotropic limit,
these models are equivalent to one-dimensional (1D)
quantum Hamiltonians, based on the TL generators.
In this paper, we suggest to study TL Hamiltonians directly,
with no necessary connection to a 2D statistical model. Various realizations
of the TL algebra will lead to distinct applications.

The TL algebra can be realized by spin-$\half$ variables.
For a loop weight $n=2$, TL Hamiltonians describe spin-$\half$ chains with isotropic
exchange interactions ${\bf S}_j \cdot {\bf S}_{j'}$. 
For nearest-neighbor interactions, one gets the Heisenberg chain~\cite{XXZ}, which is
gapless, and has a Bethe Ansatz solution. When frustrating next-nearest neighbor interactions 
are introduced (with a sufficiently strong coupling constant), the system enters
the Majumdar-Ghosh (MG) gapped phase~\cite{MG69}, characterized
by spontaneous dimerization and incommensurate correlations~\cite{MGall}.
One can extend the construction of a Hamiltonian with exact dimerized eigenstates to
a TL algebra with general loop weight $n$~\cite{qMG}. 
The new exact results we present on this model then apply to anisotropic
frustrated spin chains.

Another realization of the TL algebra, with important physical implications,
comes from the construction of trial wavefunctions for the Fractional Quantum Hall Effect (FQHE). 
The principle for this construction is to find an $\cal N$-point function 
with a prescribed behavior when several points come close to one another.
In~\cite{RR}, it was discovered that, for a specific class of wavefunctions,
the solution is provided by the correlation functions of the $\Zbb_k$-parafermionic 
Conformal Field Theories (CFTs), and that the 
elementary excitations (anyons) above the ground state obey the fusion rules of these CFTs.
For $k=3$, the resulting wavefunctions correctly describe the FQHE at filling fraction
$\nu=12/5$~\cite{FQHE}. It is important to notice that the fusion rules for the $\Zbb_k$
parafermions are encoded in the RSOS representation of the TL algebra. This fact has been 
exploited in recent works~\cite{anyons1,anyons2}, in which 1D chains of anyons 
with local TL interactions are studied. In particular, when two- and three-site interactions
are present, the phase diagram contains an MG-like gapped phase, separating two critical
phases. These anyonic chains are a particular case of the model presented in this paper.

Our approach consists in writing a 1D TL Hamiltonian
which combines two- and three-site interactions,
and studying its phase diagram.
This model contains as special cases: (i) the critical {\rm XXZ} 
spin chain; (ii) the $q$-deformed MG chain~\cite{qMG}; (iii) a new critical,
integrable point. The main results we present in this paper are the following.
For a loop weight $n \in [n^*,2]$, where $n^* \simeq 1.5$,
an MG-like gapped phase is present.
The critical line separating the {\rm XXZ} and MG phases is in the universality 
class of the {\it dilute} $\On$ model. We locate this transition line, and identify the coupling
constant in the 1D TL Hamiltonian which plays the role of temperature in the $\On$ model.
Finally, we identify a new integrable point determining a whole critical phase,
and solve it by Bethe Ansatz. Since this critical phase is separated from the {\rm XXZ} phase
by a gapped phase, the exact results from Bethe Ansatz are intrinsically non-perturbative.

\section{The model}

\begin{figure}
  \begin{center}
    \includegraphics{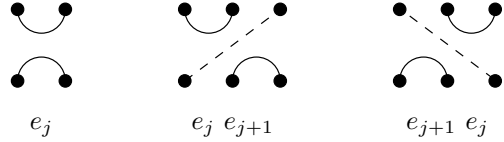}
    \caption{Action of the TL operators. Dots represent spin-$\half$ variables, and operators act upwards.
      Two spins connected by a dotted (resp. full) line are in the same state (resp. form a singlet).}
    \label{fig:tl}
  \end{center}
\end{figure}

We consider the following Hamiltonian on a chain of $2N$ sites, based on the TL generators~$e_j$
(see definition below):
\begin{equation} \label{eq:H}
  H = K_1 \sum_{j=1}^{2N} e_j + K_2 \sum_{j=1}^{2N} (e_j e_{j+1} +e_{j+1} e_j)\,.
\end{equation}
In the spin-$\half$ representation, the operator $e_j$ is a nearest-neighbor interaction, that projects the spins ${\bf S}_j, {\bf S}_{j+1}$ onto the singlet state. In this language, the quadratic terms in~\eref{eq:H} are then next-nearest-neighbor interactions (see Fig.~\ref{fig:tl}). Let us remind the definition of the TL algebra, and use the spin-$\half$ representation to write~\eref{eq:H} as a combination of projectors.

\begin{figure}
  \begin{center}
    \includegraphics{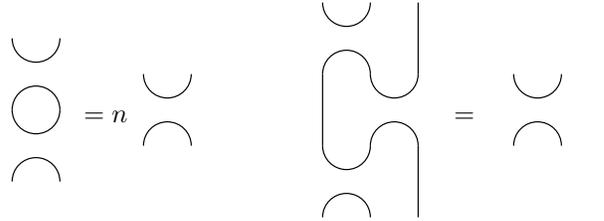}
    \caption{The Temperley-Lieb algebraic relations~\eref{eq:tl-def}.}
    \label{fig:tl-def}
  \end{center}
\end{figure}

The TL algebra with loop weight $n$ is generated by the operators $e_j$ ($j=1, \dots, 2N$), which satisfy the relations (see Fig.~\ref{fig:tl-def}):
\begin{equation} \label{eq:tl-def}
  \begin{array}{rcl}
    e_j^2 &=& n \ e_j  \\
    e_j \ e_{j \pm 1} \ e_j &=& e_j \\
    e_j \ e_{j'} &=& e_{j'} \ e_j \quad \hbox{for $|j-j'|>1$} \,.
  \end{array}
\end{equation}
The structure of the spin representation is encoded in the $\Uq$ quantum algebra~\cite{Pasquier-Saleur90}, generated by $S^z$ and $S^{\pm}$, with the $q$-deformed commutation relations:
\begin{equation}
  [S^+,S^-]=[2S^z]_q, \quad [S^z,S^{\pm}]=\pm S^{\pm}\,,
\end{equation}
where $[x]_q = (q^x-q^{-x})/(q-q^{-1})$. For $q=1$, one recovers the usual $\SU(2)$ relations.
Let us denote by $V(\ell)$ the spin-$\ell$ representation. 
The spin-$\half$ representation is simply obtained from the Pauli matrices $S^{x,y,z} = \half \sigma^{x,y,z}$, and the vector space for a spin-$\half$ chain is the tensor product $V(\half)^{\otimes 2N}$. 
Like in $\SU(2)$, the $V(\ell)$ follow the decompositions: $V(\ell) \otimes V(\ell') = V(|\ell-\ell'|) \oplus \dots \oplus V(\ell+\ell')$. We call $P_{j_1 \dots j_r}^{(\ell)}$ the projector of spins ${\bf S}_{j_1}, \dots, {\bf S}_{j_r}$ onto $V(\ell)$ according to these decompositions.
It can be shown that the rescaled projectors $e_j \equiv (q+q^{-1}) P_{j,j+1}^{(0)}$
satisfy the TL relations~\eref{eq:tl-def} with loop weight $n=q+q^{-1}$. 
Moreover, $P_{j,j+1,j+2}^{(3/2)}$ can be expressed in terms of the $e_j$:
\begin{equation*}
  P_{j,j+1,j+2}^{(3/2)} = 1 + \frac{(e_j e_{j+1} + e_{j+1} e_j) -n (e_j+e_{j+1})}{n^2-1}\,,
\end{equation*}
and thus we can write~\eref{eq:H} as:
\begin{equation*}
  H = -n \cos \theta \sum_{j=1}^{2N} P_{j,j+1}^{(0)} 
  - (n^2-1) \sin \theta \sum_{j=1}^{2N}(P_{j,j+1,j+2}^{(3/2)}-1)\,.
\end{equation*}
The original parameters are given in terms of $\theta$ and $n$ by:
\begin{equation}
  \begin{array}{rcl}
    K_1 &=& 2n \sin \theta - \cos \theta \\
    K_2 &=& -\sin \theta \,.
  \end{array}
\end{equation}

As suggested in Fig.~\ref{fig:tl-def}, the $e_j$ can be represented on a vector space of connectivity diagrams. In this {\it loop representation}, each basis state is labelled by a non-intersecting pairing of the $2N$ sites on the chain, and each pairing represents a $\Uq$ singlet state. The sector with total spin $l$ corresponds to states where $2l$ sites remain unpaired.
We denote $\Hloop$ the Hamiltonian~\eref{eq:H} in this representation.

When $q$ is a root of unity ($q={\rm e}^{i \pi/(p+1)}$, $p \geq 3$ integer), another representation of the TL algebra~\eref{eq:tl-def} is available: the RSOS or height representation~\cite{RSOS}. The site $j$ carries a height variable $h_j \in \{1, \dots, p\}$, such that $|h_j-h_{j+1}|=1$. The action of $e_j$ on the basis states is defined as:
\begin{equation}\label{eq:ej-rsos}
    e_j |h_1 \dots h_{2N}\rangle = 
    \delta_{h_{j-1},h_{j+1}}
    \sum_{h'_j} \frac{\sqrt{[h_j]_q [h'_j]_q}}{[h_{j+1}]_q}
    \ |h_1 \dots h'_j \dots h_{2N}\rangle\,, 
\end{equation}
where the sum is on valid height configurations.
These operators satisfy the TL relations~\eref{eq:tl-def} with loop weight $n=2 \cos [\pi/(p+1)]$. 
We denote $\Hrsos$ the Hamiltonian~\eref{eq:H} in this representation.
Unlike the spin and loop representations above, the $e_j$ in \eref{eq:ej-rsos} are real symmetric matrices, and $\Hrsos$ is Hermitian.
For $p=3$ (critical Ising model), additional relations exist~\cite{Saleur-Altschuler91}: $1-\sqrt{2}(e_j+e_{j+1})+(e_j e_{j+1} + e_{j+1} e_j)=0$, and hence $\Hrsos(\theta)$ is equivalent to $\pm \Hrsos(0)$ for any value of $\theta$.

There are three remarkable values of $\theta$.
For $\theta=0$, we get the XXZ Hamiltonian~\cite{XXZ}:
\begin{equation} \label{eq:Hxxz}
  \Hxxz = 2 \sum_{j=1}^{2N} \left[ S_j^x S_{j+1}^x + S_j^y S_{j+1}^y - \Delta \left(S_j^z S_{j+1}^z-\frac{1}{4}\right) \right]\,,
\end{equation}
where $\Delta = -\half (q+q^{-1})$. 
For $\theta=-\pi/2$, we get the $q$-deformed MG Hamiltonian~\cite{qMG}:
\begin{equation} \label{eq:Hmg}
  \Hmg = (n^2-1) \sum_{j=1}^{2N} (P_{j,j+1,j+2}^{(3/2)}-1)\,.
\end{equation}
For the special value $\thetaint = {\rm Arctan}(1/n) -\pi$, the Hamiltonian is integrable
(see Section~\ref{sec:int-6v}).

\section{Phase diagram}

\begin{figure}
  \begin{center}
    \includegraphics{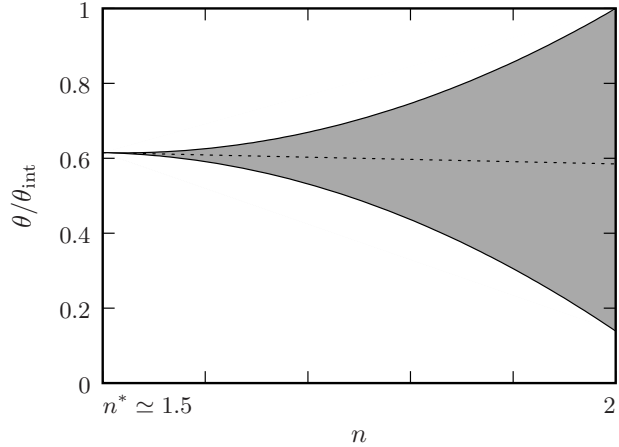}
    \caption{Sketch of the phase diagram of the Hamiltonian~\eref{eq:H}. Solid lines represent 
      transitions between the phases, and the dotted line is the MG value. The shaded area is the 
      gapped phase.}
    \label{fig:diag}
  \end{center}
\end{figure}

For a loop weight $n$ greater than $n^* \simeq 1.5$, the Hamiltonian~\eref{eq:H} has
a gapped phase of the same nature as the MG phase~\cite{MG69}.
We describe the phase diagram in the interval $\thetaint < \theta < 0$.
We use the notation:
\begin{equation}
  n = q+q^{-1} = 2 \cos \frac{\pi}{p+1}\,, \quad p \geq 1\,,
\end{equation}
where $p$ can take real values. We obtain numerically the lowest eigenvalues of $\Hloop$ and $\Hrsos$ with the Arnoldi and Lanczos algorithm, respectively. We diagonalize blocks of fixed momentum $Q$, and $2l$ unpaired sites (for $\Hloop$), for system sizes up to $2N=24$ sites. The phase diagram consists in three phases (see Fig.~\ref{fig:diag}):

\begin{description}

\item{\bf Critical phase I (XXZ phase)} \\
  For $\theta$ close enough to $0$, the system is in a critical phase, determined by $\Hxxz$~\eref{eq:Hxxz}.
  Its continuum limit is~\cite{Alcaraz-etal88} a Coulomb gas (CG) critical theory, with central charge and 
  conformal weights:
  \begin{eqnarray} 
    c &=& 1 - \frac{6(1-g)^2}{g} \label{eq:cg-1} \\
    h_{e m} &=& \frac{1}{4} \left(\frac{e}{\sqrt{g}} + m \sqrt{g} \right)^2 
    - \frac{(1-g)^2}{4g} \label{eq:cg-2} \\
    n &=& -2 \cos \pi g\,, \label{eq:cg-3} 
  \end{eqnarray}
  with $0 \leq g \leq 1$.
  The integers $e,m$ are the electric and magnetic charges. The scaling dimension for the sector with $2l$ unpaired sites (or $2l$-leg watermelon exponent) is $X_{2l} = 2h_{0,l}$. This theory corresponds to the dense phase of the $\On$ model. For integer $p \geq 3$, $\Hrsos(\theta=0)$ is~\cite{Huse84} a lattice realization of the minimal model ${\cal M}(p+1,p)$ of CFT~\cite{CFT}.
  
\item{\bf Gapped phase} \\
For generic $n \in [n^*,2]$, $\Hloop(\theta)$ has a gapped phase around $\theta=-\pi/2$, with the following features (as in the isotropic case~\cite{MG69,MGall}): (i) the ground state is doubly-degenerate and consists of totally dimerized states; (ii) the elementary excitations (spinons) are domain walls between the two ground-state configurations, and are gapped.
In the RSOS representation for $p \geq 3$ integer, we can {\it prove} that a totally dimerized state $\psi_a(h_1,\dots,h_{2N}) = \prod_{j=1}^N \sqrt{[h_{2j-1}]_q/[a]_q} \ \delta_{h_{2j},a}$, with $a \in \{1,\dots,p\}$, saturates the lower energy bound of $\Hmg$. Indeed, since $\Hmg$ is a sum of Hermitian operators, all energies satisfy $E \geq -2N(n^2-1)$. The state $\psi_a$ is an eigenstate of $\Hmg$ with energy $-2N(n^2-1)$. The height $a$ can take $p$ values, and $\psi_a$ can be translated by one site, so the RSOS ground state of $\Hmg$ is $2p$-degenerate.

\item{\bf Critical phase II} \\
This phase is determined by $H(\thetaint)$, which is solved in Section~\ref{sec:int-6v}.

\end{description}

\begin{figure}
  \begin{center}
    \includegraphics{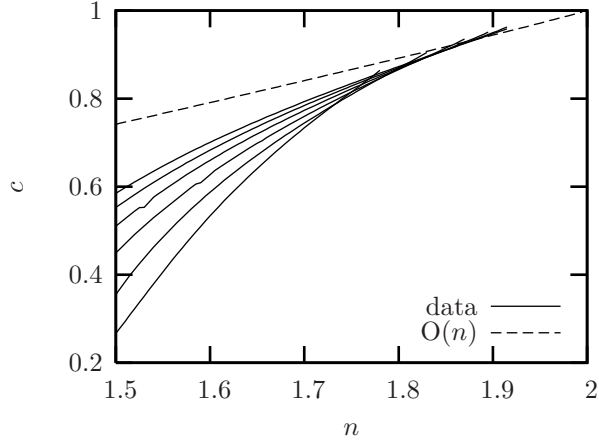}
    \caption{Effective central charge for the loop model at $\theta_c$. The solid lines are numerical results for system sizes $N=6,7$ to $N=11,12$ (from bottom to top). The dotted line shows the exact result for the dilute $\On$ model.}
    \label{fig:cc}
  \end{center}
\end{figure}

\begin{figure}
  \begin{center}
    \includegraphics{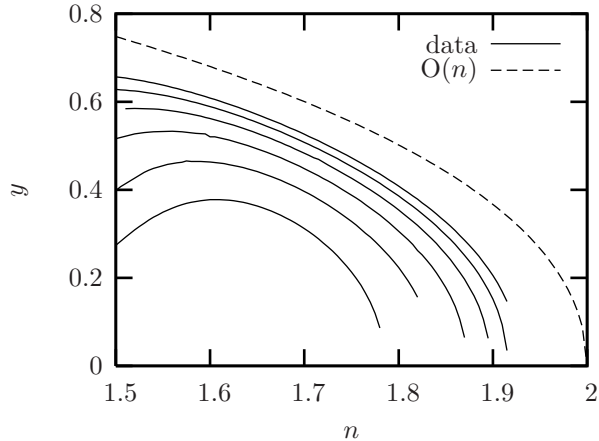}
    \caption{Scaling exponent of $\theta$ at $\theta_c$. Same legend as Fig.~\ref{fig:cc}.}
    \label{fig:y-q}
  \end{center}
\end{figure}

At the transition point $\theta_c$ between the XXZ and gapped phases, $\Hloop$ is described by a CG theory~(\ref{eq:cg-1}--\ref{eq:cg-3}), with $1 \leq g \leq 2$, corresponding to the critical dilute $\On$ model (see Fig.~\ref{fig:cc}). So, like the temperature does in the $\On$ model~\cite{Nienhuis82}, $\theta$ drives a transition between a non-critical phase and a critical phase, with a higher critical behavior at the transition.
Moreover, the scaling exponent for $\theta$ is the $\On$ thermal exponent $y_t=4/(p+2)$ (see Fig.~\ref{fig:y-q}).
For integer $p \geq 4$, the spectrum of $\Hrsos(\theta_c)$ coincides with the CFT minimal model ${\cal M}(p+2,p+1)$~\cite{CFT}, as found in~\cite{anyons2} for $p=4$.

\section{Relations to integrable six-vertex models}
\label{sec:int-6v}

From the TL generators for $n=2\cos \gamma$ ($0 \leq \gamma \leq \pi/2$), one builds an $\Rc$-matrix, acting on sites $j,j+1$: $\Rc_j(u) = \sin(\gamma-u) \ 1 + \sin u \ e_j$, which satisfies the Yang-Baxter equation:
$\Rc_j(u) \Rc_{j+1}(u+v) \Rc_j(v) = \Rc_{j+1}(v) \Rc_j(u+v) \Rc_{j+1}(u)$.
In the spin-$\half$ representation, $\Rc_j(u)$ encodes the integrable six-vertex (6V) model.
Consider the staggered 6V model on a square lattice of width $2N$ sites, defined by the spectral parameters $(u, u+\alpha, u, u+\alpha \dots)$ in the horizontal direction, and $(0, \alpha, 0, \alpha \dots)$ in the vertical direction. Let $T_\alpha(u)$ be the two-row transfer matrix. In the anisotropic limit $u \to 0$, the Hamiltonian is: ${\cal H}_\alpha \equiv (\rho \sin \gamma / \sin \alpha) \, T_\alpha(0)^{-1} T'_\alpha(0)$, where $\rho = \sin(\gamma-\alpha) \sin(\gamma+\alpha)$. We get:
\begin{eqnarray}
  {\cal H}_\alpha = \sum_{j=1}^{2N} &\Bigg[& 
    \frac{2 \rho}{\sin \alpha} \, e_j + \cos \gamma \sin \alpha 
    \, (e_j e_{j+1} + e_{j+1} e_j) \nonumber \\
    &-& \sin \gamma \cos \alpha \, (-1)^j (e_j e_{j+1} - e_{j+1} e_j)
  \Bigg]\,. \label{eq:H-stag}
\end{eqnarray}

From now on, we focus on the case $\alpha = \pi/2$: the Hamiltonian~\eref{eq:H-stag} then has the form~\eref{eq:H}, with $K_2 = -K_1/n$. The loop model defined by $T_{\pi/2}(u)$ is equivalent to the critical antiferromagnetic (AF) Potts model~\cite{Baxter-PAF}, but the sign of ${\cal H}_{\pi/2}$ is such that the its lowest energies correspond to the {\it lowest} eigenvalues of $T_{\pi/2}(u)$. Therefore, the scaling behavior of ${\cal H}_{\pi/2}$ is different from the AF Potts model studied in~\cite{Baxter-PAF,IJS08}.
For periodic boundary conditions, the Bethe Ansatz Equations (BAE) and energy for $r$ particles read:
\begin{equation}
  \begin{array}{c}
    {\displaystyle \left[ \frac{\sinh (\alpha_j-i\gamma)}{\sinh (\alpha_j+i\gamma)} \right]^N
      = - \prod_{l=1}^r \frac{\sinh (\frac{\alpha_j-\alpha_l}{2} -i\gamma)}{\sinh (\frac{\alpha_j-\alpha_l}{2} +i\gamma)} } \\
    {\displaystyle E = 2N \cos 2 \gamma - \sum_{j=1}^r \frac{2 \sin^2 2\gamma}{\cosh 2 \alpha_j - \cos 2 \gamma}\,.}
  \end{array}
\end{equation}
The ground state has $N/2$ Bethe roots on each of the lines ${\rm Im} \ \alpha_j = 0$ and ${\rm Im} \ \alpha_j = \pi$. Since each of these two Fermi seas can be excited {\it independently} by creating holes, the central charge of the staggered 6V model is $c=2$. Solving the BAE by Fourier transform in the large-$N$ limit, we obtain the spectrum of conformal dimensions:
\begin{eqnarray}
  \Delta_{e m,\et \mt} &=& \frac{1}{8} \left( \frac{e}{\sqrt{2g'}} + m \sqrt{2g'} \right)^2
  + \frac{1}{8} (\et+\mt)^2 \\
  \bar{\Delta}_{e m,\et \mt} &=& \frac{1}{8} \left( \frac{e}{\sqrt{2g'}} - m \sqrt{2g'} \right)^2
  + \frac{1}{8} (\et-\mt)^2 \\
  g' &=& (\pi-2\gamma)/(2\pi) \,,
\end{eqnarray}
where $e,\et$ (resp. $m,\mt$) are integers with the same parity. To achieve the equivalence between the 6V and loop models, we introduce a seam with phase $\pm \pi e_0$, where  $e_0= \gamma/\pi$. The effective central charge is then $c_{\rm eff} = 2 - 6 e_0^2/g'$. The expressions for $c_{\rm eff}$ and $\Delta_{e m,\et \mt}$ suggest that the field theory for the staggered 6V model consists of one free compact boson and two free Majorana fermions, decoupled in the bulk, but coupled through the boundary conditions.

\section{Conclusion}

We have defined a Hamiltonian $H(\theta)$ based on the Temperley-Lieb algebra, which contains, as
particular points, the XXZ and $q$-deformed Majumdar-Ghosh spin chains. At the transition $\theta_c$
between the two corresponding phases, we have given numerical evidence that the spectrum of $H(\theta_c)$
is equivalent to that of the dilute $\On$ model. For the integrable point $\thetaint$ beyond the MG phase, 
we have used the BAE to find the effective degrees of freedom in the continuum limit.
From these results, it is tempting to suppose that $H(\theta_c)$ is also the anisotropic limit of some
integrable transfer matrix, but this is not proven. A related question is the exact determination of
the limiting value $n^*$, where the gapped phase ceases to exist.

\ack{We thank Fabien Alet for instructive discussions and help on numerics. YI also thanks Fabian Essler for help with the literature. The work of JLJ was supported by the ANR; the work of HS by the ANR and the ESF Network INSTANS.}

\end{document}